\documentclass[12pt,draft,notref,notcite]{article}

\usepackage{color,longtable}
\usepackage{amssymb,graphicx}

\newcommand{\be}{\begin{eqnarray}}
\newcommand{\ee}{\end{eqnarray}}

\usepackage{amsfonts,amsmath}
\usepackage{latexsym}

\def\cH{{\cal H}}

\def\a{{\alpha}}
\def\b{{\beta}}

\def\E{E_{10}}
\def\K{K(E_{10})}
\def\KE{E_{10}/K(E_{10})}

\begin{document}

\begin{center}

{\bf \Large Chaos and Symmetry in String Cosmology}\footnote{Invited talk at the 11th Marcel Grossmann Meeting on Recent Developments in General Relativity, Berlin, Germany, 23-29 July 2006.}

\bigskip

Thibault Damour

\medskip

{\sl Institut des Hautes Etudes Scientifiques, 35 route de Chartres,  \\ F-91440 Bures-sur-Yvette, France}
\end{center}

\vspace{1cm}

\begin{minipage}{12cm}
\textbf{Abstract:}
We review the recently discovered interplay between chaos and symmetry in the general inhomogeneous solution of many string-related Einstein-matter systems in the vicinity of a cosmological singularity. The Belinsky-Khalatnikov-Lifshitz-type chaotic behaviour is found, for many Einstein-matter models (notably those related to the low-energy limit of superstring theory and $M$-theory), to be connected with certain (infinite-dimensional) hyperbolic Kac-Moody algebras. In particular, the billiard chambers describing the asymptotic  cosmological behaviour of pure Einstein gravity in spacetime dimension $d+1$, or the metric-three-form system of $11$-dimensional supergravity, are found to be identical to the Weyl chambers of the Lorentzian Kac-Moody algebras $AE_d$, or $E_{10}$, respectively. This suggests that these Kac-Moody algebras are hidden symmetries of the corresponding models. There even exists some evidence of a hidden equivalence between the general solution of the Einstein-three-form system and a null geodesic in the infinite dimensional coset space $E_{10} / K (E_{10})$, where $K (E_{10})$ is the maximal compact subgroup of $E_{10}$.
\end{minipage}

\vspace{1cm}

\section {Introduction}

We wish to review a recently discovered intriguing connection between two, seemingly antagonistic, structures present in pure Einstein gravity, and in some, string-theory motivated, Einstein-matter systems.

On the one hand, Belinsky, Khalatnikov and Lifshitz (BKL) \cite{Belinsky:1970ew} discovered that the asymptotic behaviour of the general solution of the $(3+1)$-dimensional Einstein's equations, in the vicinity of a cosmological singularity, exhibited a {\it chaotic structure}: they showed that, because of non-linearities in Einstein equations, the generic, inhomogeneous solution behaves as a chaotic \cite{Khalatnikov:1984ck} sequence of ``generalized Kasner solutions''. They then showed that pure gravity in $4+1$ dimensions exhibits a similar chaotic structure \cite{BK73}. The extension of the BKL analysis to pure gravity in higher dimensions was addressed in \cite{Demaret:1986su,Demaret:1986ys}. A surprising result was found: while the general behaviour of the vacuum Einstein solutions remains ``chaotic'' (i.e. ``oscillatory'') for spacetime dimensions $D \leq 10$, it ceases to be so for spacetime dimensions $D \geq 11$, where it becomes monotonic and Kasner-like.

On the other hand, a {\it symmetry structure} was found to be present (often in a hidden form) in the ``dimensional reductions'' of Einstein's gravity. The paradigm of this hidden symmetry structure is the continuous $SL(2,{\mathbb R})_E$ symmetry group found long ago by Ehlers \cite{E62} for $D=4$ Einstein gravity in the presence of {\it one Killing vector}. When {\it two commuting Killing vectors} are present, the finite-dimensional Ehlers group $SL(2,{\mathbb R})$ was  found, through, notably, the work of Matzner and Misner, of Geroch, of Julia, of Breitenlohner and Maison, and of Belinsky and Zakharov, to be promoted to an infinite-dimensional symmetry group, which can be identified to the affine Kac-Moody extension of $SL(2,{\mathbb R})$ (see \cite{Nicolai:1991tt} for references and a review). A similar pattern was found to take place in supergravity theories, and notably in the ``maximal'' supergravity theory which lives in $D=11$. In that case, it was remarkably found \cite{Cremmer:1979up} that the successive toroidal dimensional reductions of $D=11$ supergravity to lower dimensions (in the presence of an increasing number of commuting Killing vectors) admitted a correspondingly larger and larger hidden symmetry group $E_n ({\mathbb R})$, where $n$ denotes the number of Killing vectors, and $E_n$ the (extended) sequence of the groups belonging to the {\it exceptional series} in the Cartan-Killing classification of finite-dimensional simple Lie groups. The latter series culminates into the last finite-dimensional exceptional group $E_8$ in the case where one has 8 Killing vectors, i.e. in the case of toroidal compactification (on $T^8$) down to a $D' = 3$ dimensionally reduced theory. However, as first conceived by Julia \cite{Julia:1982gx}, this symmetry-increasing pattern is expected to continue to the (infinite-dimensional) affine Kac-Moody group $E_9$ for the reduction of supergravity down to $D'=2$ (9 Killing vectors). This was indeed explicitly proven later \cite{Nicolai:1987kz}. It was also mentionned by Julia \cite{Julia:1982gx} that the still larger symmetry group $E_{10}$ (which is now an infinite-dimensional {\it hyperbolic} Kac-Moody group\footnote{See \cite{Kac:1990gs} for an introduction to the theory of Kac-Moody algebras.}) might arise when trying to further reduce maximal supergravity. However, this poses a serious challenge because the naive dimensional reduction to $D' = 1$, i.e. the set of $D=11$ supergravity solutions which depend only on one (time) variable is much too small to (faithfully) carry such a huge symmetry.

At this point, the two separate threads (chaos and symmetry) in our history unexpectedly merged in a sequence of works which, while extending the BKL analysis to $D=11$ supergravity, and to the $D=10$ models describing the low-energy limit of the various superstring theories \cite{Damour:2000wm,Damour:2000th}, found that behind their seeming entirely {\it chaotic}\footnote{Though $D=11$ pure gravity is monotonically Kasner-like instead of chaotic, the presence of the three-form $A$ in $D=11$ supergravity was found to reintroduce a generic chaotic, oscillatory behaviour.} BKL-type behaviour there were hints of a hidden {\it symmetry structure} \cite{hep-th/0012172,hep-th/0103094,hep-th/0207267} linked to $E_{10}$. More precisely, it was first found \cite{hep-th/0012172} that the ``billiard chamber'' (see \cite{hep-th/0212256} and below) describing the BKL-type behaviour of $D=11$ supergravity (as well as of $D=10$ IIA and IIB superstring theories) could be identified with the ``Weyl chamber'' of $E_{10}$. [The cosmological billiard chambers of type I and heterotic string theories coincides with the Weyl chamber of $BE_{10}$. We focus here on the more fundamental $E_{10}$ case.] Similarly, an examination of the case of pure gravity in spacetime dimension $D \equiv d+1$ revealed the presence of the Weyl chamber of the Lorentzian Kac-Moody algebras $AE_d$ (which are hyperbolic only when $d \leq 9$) \cite{hep-th/0103094}. Then it was shown that, up to height 30 in a ``height expansion'' related to the BKL ``gradient expansion'', the dynamics of the 11-dimensional supergravity variables could be identified with the dynamics of a null geodesic on the infinite-dimensional coset space $E_{10} / K (E_{10})$, where $K(E_{10})$ is the maximal compact subgroup of $E_{10}$ \cite{hep-th/0207267}. This led to the conjecture that there exists a hidden equivalence between $D=11$ supergravity and null geodesic motion on $E_{10} / K(E_{10})$. This conjecture can be generalized to other gravity models and other coset spaces \cite{hep-th/0212256}. In particular, usual $(3+1)$-dimensional gravity might be equivalent to null geodesic motion on $AE_3 / K(AE_3)$. If these conjectures were true, they would mean that the infinite-dimensional numerator Kac-Moody groups, $E_{10}$ or $AE_3$, are hidden continuous ``symmetries'' of the respective field equations (supergravity$_{11}$ or gravity$_4$) which transform solutions into (new) solutions. In the case of $D=4$ gravity, this would represent a huge generalization\footnote{This is a generalization in two separate senses: (i) the conjectured $AE_3$ symmetry would apply in absence of any Killing field, and (ii) $AE_3$ is the {\it hyperbolic} Kac-Moody group canonically associated with $A_1 = SL(2)$ (while the ``Geroch group'' is the affine Kac-Moody group associated with $A_1 = SL(2)$).} of the Ehlers $SL(2,{\mathbb R})$ symmetry group.

The present paper is organized as follows. In Section~2 we summarize the Hamiltonian ``cosmological billiard'' approach to BKL behaviour (mainly drawing from \cite{hep-th/0212256}). See \cite{gr-qc/0702141} and the contribution of C.~Uggla to these proceedings for a comparison between this Hamiltonian cosmological billiard approach and a ``Hubble-normalized'' ``dynamical systems picture''. In Section~3 we sketch the ``correspondence'' between a gravity model (e.g. $D=11$ supergravity) and a coset geodesic dynamics (e.g. geodesic motion on $E_{10} / K(E_{10})$). For more details on the construction and structure of those Kac-Moody coset models see the contribution of A.~Kleinschmidt and H.~Nicolai to these proceedings. Finally, Section~4 offers some (speculative) conclusions.

\section{Cosmological billiards}

Let us start by summarizing the BKL-type analysis of the ``near spacelike singularity limit'', that is, of the asymptotic behaviour of the metric $g_{\mu\nu}(t, \textbf{x})$, together with the other fields (such as the 3-form $A_{\mu \nu\lambda} (t, \textbf{x})$  in supergravity), near a singular hypersurface. The basic idea is that, near a spacelike singularity, the time derivatives are expected to dominate over spatial derivatives. More precisely, BKL found that spatial derivatives introduce terms in the equations of motion for the metric which are similars to the ``walls'' of a billiard table \cite{Belinsky:1970ew}. To see this, it is convenient \cite{hep-th/0212256} to decompose the $D$-dimensional metric $g_{\mu\nu}$ into non-dynamical (lapse $N$, and shift $N^i$, here set to zero) and dynamical ($e^{- 2 \beta^a}$, $ \theta^{a}_i)$ components. They are defined so that the line element reads 
\begin{equation}
\label{eq1}
ds^2 = - N^2 dt^2 + \sum_{a = 1}^{d} 
e^{-2 \beta^{a}} \theta^a_i \theta^a_j  dx^i dx^j. 
\end{equation}
Here $d \equiv D- 1$ denotes the spatial dimension ($d = 10$ for SUGRA$_{11}$, and $d = 9$ for string theory), $e^{- 2 \beta^a}$ represent (in an Iwasawa decomposition) the ``diagonal'' components of the spatial metric $g_{ij}$, while the ``off diagonal'' components are represented by the $\theta^{a}_i$, defined to be upper triangular matrices with 1's on the diagonal (so that, in particular, $\det \theta = 1$).

The Hamiltonian constraint, at a given spatial point, reads (with $ \tilde{N} \equiv N / \sqrt{\det g_{ij}}$ denoting the ``rescaled lapse'')
\begin{eqnarray}
\label{eq2}
&&\cH(\b^a, \pi_{a},P,Q) \nonumber \\
&= &\tilde{N} \left[\frac12 \ G^{ab} \pi_a \pi_b  +
\sum_A c_A (Q,P,\partial\b,\partial^2 \b, \partial Q)\exp\big(- 2 w_A (\b)\big)\right] \, .
\end{eqnarray}
Here $\pi_{a}$ (with $a = 1, ... ,d$) denote the canonical momenta conjugate to the ``logarithmic scale factors'' $\beta^a$, while $Q$ denote the remaining configuration variables ($\theta^a_i$, 3-form components $A_{ijk} (t, \textbf{x})$ in supergravity), and $P$ their canonically conjugate momenta ($P^i_a, \pi^{ijk}$). The symbol $\partial$ denotes {\em spatial}  derivatives.  The (inverse) metric $G^{ab}$ in Eq.~(\ref{eq2}) is the DeWitt ``superspace'' metric induced on the $\beta$'$s$ by the Einstein-Hilbert action. It endows the  $d$-dimensional\footnote{10 dimensional for SUGRA$_{11}$; but the various superstring theories also lead to a 10 dimensional Lorentz space   because one must add the (positive) kinetic term of the dilaton $\varphi \equiv \beta^{10} $ to the 9-dimensional DeWitt metric corresponding to the 9 spatial dimensions.} $\beta$ space with a Lorentzian structure
$ G_{ab} \, \dot{\beta}^a \dot{\beta}^b$. 

One of the crucial features of Eq.~(\ref{eq2}) is the appearance of Toda-like exponential potential terms 
$\propto \exp (- 2 w_A(\beta))$, where the $w_A(\beta)$ are {\em linear forms} in the logarithmic scale factors: $w_A(\beta) \equiv w_{Aa} \, \beta^a$. The range of labels $A$ and the specific ``wall forms'' $w_A(\beta)$ that appear depend on the considered model. For instance, in SUGRA$_{11}$ there appear: ``symmetry wall forms'' $w_{ab}^S (\beta) \equiv \beta^b - \beta^a$ (with $a < b$), ``gravitational wall forms'' $w_{abc}^g (\beta) \equiv 2\beta^a + \underset{e \neq a,b, c}{\sum} \,  \beta^e$ ($a \neq b$, $b \neq c$, $c \neq a$), ``electric 3-form wall forms'', $e_{abc} (\beta )
\equiv \beta^{a} + \beta^{b} + \beta^{c}$ ($a \neq b$, $b\neq c$, $c\neq a$), and ``magnetic 3-form wall forms'', $m_{a_1 .... a_6} \equiv \beta^{a_1} + \beta^{a_2} + ... + \beta^{a_6}$ (with indices all different).

One then finds that the near-spacelike-singularity limit amounts to considering the {\em large $\beta$ limit} in Eq.(2). In this limit a crucial role is played by the linear forms $w_A ( \beta)$ appearing in the ``exponential walls''. Actually, these walls enter in successive ``layers''. A first layer consists of a subset of all the walls called the {\em dominant walls} $w_i (\beta)$. The effect of these dynamically dominant walls is to confine the motion in $\beta$-space to a {\em fundamental billiard chamber} defined by the inequalities $w_i (\beta) \geqslant 0$. In the case of SUGRA$_{11}$, one finds that there are 10 dominant walls: 9 of them are the symmetry walls  $w_{12}^{S}(\beta), w_{23}^{S}(\beta), ... , w_{910}^{S}(\beta), $ and the 10th is an electric 3-form wall  $e_{123} (\beta) = \beta^1 + \beta^2 + \beta^3$. As noticed in \cite {hep-th/0012172} a remarkable fact is that the fundamental cosmological billiard chamber of SUGRA$_{11}$ (as well as type-II string theories) is the {\em Weyl chamber} of the hyperbolic Kac-Moody algebra $\E$. More precisely, the 10 dynamically dominant wall forms $\big\{w_{12}^{S}(\beta), w_{23}^{S}(\beta), ... , w_{910}^{S}(\beta), e_{123} (\beta)\big\}$  can be identified with the 10 {\em simple roots}    $ \{ \alpha_1(h), \alpha_2(h), ... , \alpha_{10}(h)\}$ of $E_{10}$. Here $h$ parametrizes a generic element of a Cartan subalgebra (CSA) of $E_{10}$ . [Let us also note that for Heterotic and type-I string theories the cosmological billiard is the Weyl chamber of another rank-10 hyperbolic Kac-Moody algebra, namely $B E_{10}$]. In the Dynkin diagram of $E_{10}$, Fig. 1, the 9 ``horizontal'' nodes correspond to the 9 symmetry walls, while the characteristic ``exceptional'' node sticking out ``vertically'' corresponds to the electric 3-form wall $e_{123} = \beta^1 + \beta^2 + \beta^3$.  [The fact that this  node stems from the {\em 3rd} horizontal node is then seen to be directly related to the presence of the {\it 3-form} $A_{\mu\nu\lambda}$, with electric kinetic energy $ \propto g^{i \ell} g^{jm} g^{kn} \dot{A}_{ijk}\dot{A}_{\ell m n}$].

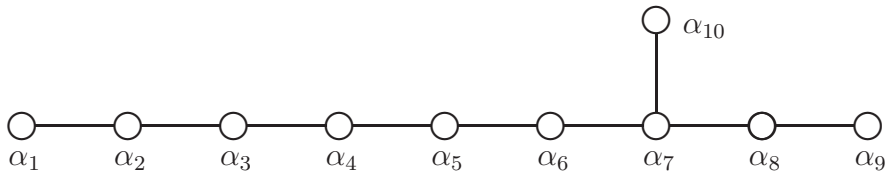
\begin{figure}[h]
\begin{center}
\scalebox{1}{
\begin{picture}(340,60)
\put(5,-5){$\alpha_1$}
\put(45,-5){$\alpha_2$}
\put(85,-5){$\alpha_3$}
\put(125,-5){$\alpha_4$}
\put(165,-5){$\alpha_5$}
\put(205,-5){$\alpha_6$}
\put(245,-5){$\alpha_7$}
\put(285,-5){$\alpha_8$}
\put(325,-5){$\alpha_9$}
\put(260,45){$\alpha_{10}$}
\thicklines
\multiput(10,10)(40,0){9}{\circle{10}}
\multiput(15,10)(40,0){8}{\line(1,0){30}}
\put(250,50){\circle{10}} \put(250,15){\line(0,1){30}}
\put(290,10){\circle{10}}
\end{picture}}
\caption{\label{e10dynk}\sl Dynkin diagram of $E_{10}$.}
\end{center}
\end{figure}

The  appearance of $\E$ in the BKL behaviour of SUGRA$_{11}$ revived the old suggestion of  \cite{Julia:1982gx} about the possible role of $\E$ in a {\em one-dimensional reduction } of SUGRA$_{11}$. A posteriori, one can view the BKL behaviour as a kind of spontaneous reduction to one dimension (time) of a multidimensional theory. Note, however, that we are always discussing generic {\em inhomogeneous} 11-dimensional solutions, but that we examine them in the near-spacelike-singularity limit where the spatial derivatives are sub-dominant: $\partial_x \ll \partial_t$. Note also that the discrete $E_{10} (\mathbb{ Z})$ was proposed as a $U$-duality group of the  full $(T^{10})$ spatial toroidal compactification of $M$-theory by Hull and Townsend \cite{hep-th/9410167}.

\section{Gravity/Coset correspondence}

Refs \cite{hep-th/0207267,hep-th/0410245} went beyond the leading-order BKL analysis just recalled by including the first three ``layers'' of spatial-gradient-related sub-dominant walls $\propto \exp (- 2 w_A (\beta))$ in Eq.(\ref{eq2}).  The relative importance of these sub-dominant walls, which modify the leading billiard dynamics defined by the 10 dominant walls $w_i(\beta)$, can be ordered   by means of an expansion which counts how many dominant wall forms $w_i(\beta)$ are contained in the exponents of the sub-dominant wall forms $w_A(\beta)$, associated  to {\em higher spatial gradients}. By mapping the dominant gravity wall forms $w_i(\beta)$ onto the corresponding $\E$ simple roots $\a_i (h), i = 1, ..., 10,$ the just described BKL-type {\em gradient expansion} becomes mapped onto a Lie-algebraic {\it height expansion} in the roots of $\E$.  It was remarkably found that, up to height 30 (i.e. up to small corrections to the billiard dynamics associated to the product of 30 leading walls $e^{- 2 w_i (\beta) }$), the SUGRA$_{11}$ dynamics for $g_{\mu\nu} (t, \textbf{x})$, $A_{\mu\nu\lambda} (t, \textbf{x})$, considered  at some given spatial point $\textbf{x}_0$, could be identified to the geodesic dynamics of a {\em massless particle} moving on the (infinite-dimensional) coset space $\KE$. Note the ``holographic'' nature of this correspondence between an 11-dimensional dynamics on one side, and a 1-dimensional one on the other side.

A point on the coset space $E_{10}(\mathbb{R}) / K(E_{10}(\mathbb{R}))$ is coordinatized by a time-dependent (but spatially independent) element of the $E_{10}(\mathbb{R})$ group of the (Iwasawa) form: $g(t) = \exp h(t) \exp \nu (t)$. Here, $ h(t) = \beta_{\textrm{coset}}^a (t) H_a$ belongs to the 10-dimensional CSA of $E_{10}$, while $\nu (t) = \sum_{\alpha > 0} \nu^\alpha (t) E_\alpha$ belongs to a Borel subalgebra of $E_{10}$ and has an infinite number of components labelled by a {\em positive root} $\alpha $ of $E_{10}$.  The (null) geodesic action over the coset space $\KE$ takes the simple form
\begin{equation}
\label{eq3}
S_{E_{10}/K(E_{10})} = \int \frac{dt}{n(t)} (v^{\textrm{sym}} | v^{\textrm{sym}})
\end{equation}
where $v^{\textrm{sym}} \equiv \frac{1}{2} (v + v^T)$ is the ``symmetric''\footnote{Here the transpose operation $T$ denotes the negative of the Chevalley involution $\omega$ defining the real form  $E_{10(10)}$ of $E_{10}$. It is such that the elements $k$ of the Lie sub-algebra of $\K$ are ``$T$-antisymmetric'': $k^T = - k $, which is equivalent to them being fixed under $\omega : \omega(k)= + \, \omega(k)$.} part of the ``velocity''  $v \equiv (dg/dt) g^{-1}$ of a group element $g(t)$ running over $E_{10}(\mathbb{R})$.

The correspondence between the gravity, Eq.~(\ref{eq2}), and coset, Eq.~(\ref{eq3}), dynamics is best exhibited by decomposing (the Lie algebra of) $E_{10}$ with respect to (the Lie algebra of) the $GL(10)$ subgroup defined by the horizontal line in the Dynkin diagram of $E_{10}$. This allows one to grade the various components of $g(t)$ by their $GL(10)$ level $\ell$. One finds that, at the $\ell = 0$ level, $g(t)$ is parametrized by the Cartan coordinates $\beta_{\rm coset}^a (t)$ together with a unimodular upper triangular zehnbein $\theta_{{\rm coset} \, i}^a (t)$. At level $\ell = 1$, one finds a 3-form $A_{ijk}^{\rm coset} (t)$; at level $\ell = 2$, a 6-form $A_{i_1 i_2 \ldots i_6}^{\rm coset} (t)$, and at level $\ell = 3$ a $9$-index object $A_{i_1 \mid i_2 \ldots i_9}^{\rm coset} (t)$ with Young-tableau symmetry $\{ 8,1 \}$. The coset action (\ref{eq3}) then defines a coupled set of equations of motion for $\beta_{\rm coset}^a (t)$, $\theta_{{\rm coset} \, i}^a (t)$, $A_{ijk}^{\rm coset} (t)$, $A_{i_1 \ldots i_6}^{\rm coset} (t)$, $A_{i_1 \mid i_2 \ldots i_9}^{\rm coset} (t)$. By explicit calculations, it was found that these coupled equations of motion could be identified (modulo terms corresponding to potential walls of height at least 30) to the SUGRA$_{11}$ equations of motion, considered at some given spatial point $\textbf{x}_0$. 

The {\it dictionary} between the two dynamics says essentially that: 

\noindent (0) $\beta_{\rm gravity}^a (t,\textbf{x}_0) \leftrightarrow \beta_{\rm coset}^a (t) \, , \ \theta_i^a (t, \textbf{x}_0) \leftrightarrow \theta_{{\rm coset} \, i}^a (t)$, (1) $\partial_t \, A_{ijk}^{\rm coset} (t)$ corresponds to the electric components of the 11-dimensional field strength $F_{\rm gravity}$ $= d \, A_{\rm gravity}$ in a certain frame $e^i$, (2) the conjugate momentum of $A_{i_1 \ldots i_6}^{\rm coset} (t)$ corresponds to the {\it dual} (using $\varepsilon^{i_1 i_2 \ldots i_{10}}$) of the ``magnetic'' frame components of the 4-form $F_{\rm gravity} = d \, A_{\rm gravity}$, and (3) the conjugate momentum of $A_{i_1 \mid i_2 \ldots i_9} (t)$ corresponds to the $\varepsilon^{10}$ dual (on $jk$) of the structure constants $C_{jk}^i$ of the coframe $e^i$ ($d \, e^i = \frac{1}{2} \, C_{jk}^i \, e^j \wedge e^k$).

The fact that at levels $\ell = 2$ and $\ell = 3$ the dictionary between supergravity and coset variables maps the {\it first spatial gradients} of the SUGRA variables $A_{ijk} (t,\textbf{x})$ and $g_{ij} (t,\textbf{x})$ onto (time derivatives of) coset variables suggested the conjecture \cite{hep-th/0207267} of a hidden {\it equivalence} between the two models, i.e. the existence of a dynamics-preserving map between the infinite tower of (spatially independent) coset variables $(\beta_{\rm coset}^a , \nu^{\alpha})$, together with their conjugate momenta $(\pi_a^{\rm coset} , p_{\alpha})$, and the infinite sequence of spatial Taylor coefficients $(\beta (\textbf{x}_0)$, $\pi (\textbf{x}_0)$, $Q (\textbf{x}_0)$, $P (\textbf{x}_0)$, $\partial Q (\textbf{x}_0)$, $\partial^2 \beta (\textbf{x}_0)$, $\partial^2 Q(\textbf{x}_0), \ldots , \partial^n Q (\textbf{x}_0) , \ldots)$ formally describing the dynamics of the gravity variables $(\beta (\textbf{x}) , \pi (\textbf{x}) , Q(\textbf{x})$, $P(\textbf{x}))$ around some given spatial point $\textbf{x}_0$.\footnote{One, however, expects the map between the two models to become spatially non-local for heights $\geq 30$.}

It has been possible to extend the correspondence between the two models to the inclusion of fermionic terms on both sides \cite{hep-th/0512163,hep-th/0512292,hep-th/0606105}. Moreover, Ref. \cite{hep-th/0504153} found evidence for a nice compatibility between some high-level contributions (height $-115$!) in the coset action, corresponding to {\it imaginary} roots\footnote{i.e. such that $(\alpha , \alpha) < 0$, by contrast to the ``real'' roots, $(\alpha , \alpha) = +2$, which enter the checks mentionned above.}, and $M$-theory {\it one-loop corrections} to SUGRA$_{11}$, notably the terms quartic in the curvature tensor. (See also \cite{Damour:2006ez} for a study of the compatibility of an underlying Kac-Moody symmetry with quantum corrections in various models).

\section{Conclusions and outlook}

At this stage, we are far from having a proof of the full equivalence between SUGRA$_{11}$ and the $E_{10} / K(E_{10})$ model, or of any other gravity/coset conjectured pair. The partial evidence summarized above is suggestive of the existence of some kind of hidden symmetry structure in General Relativity and Supergravity. However, it is quite possible that this hidden symmetry is present only in a way which cannot be explicitly realized at the level of the classical field equations. Indeed, the situation might be similar to that discussed in the plenary talks of Sasha Polyakov, Igor Klebanov and Eva Silverstein. In the much better understood gravity/gauge correspondence a quasi-classical, weakly curved spacetime corresponds to a strongly quantum, strongly self-interacting gauge theory state. Reciprocally, the perturbative, weakly-interacting gauge theory states correspond to non-perturbative, strongly curved gravity states. By analogy, we might expect that the simple geodesic coset motions correspond to strongly curved spacetimes, with curvatures larger than the Planck scale. [This would intuitively explain why the coset picture becomes prominent in the formal limit where one tends towards an infinite-curvature singularity.] In addition, it is possible that the equivalence between the gravity and coset models hold only at the level of the quantized models. In this respect, note that the quantum version of the null geodesic dynamics (\ref{eq3}) would be, if we neglect polarization effects\footnote{Actually, Refs~\cite{hep-th/0512163,hep-th/0512292,hep-th/0606105} indicate the need to consider a {\it spinning} massless particle, i.e. some kind of Dirac equation on $E_{10} / K(E_{10})$.} a Klein-Gordon equation\footnote{This equation, submitted to a condition of periodicity over a discrete group $E_{10} ({\mathbb Z})$, has been considered in Refs~\cite{hep-th/9903110,hep-th/0401053} in the context of quantum theories with all spatial dimensions being toroidally compactified.},
\begin{equation}
\label{eq4}
\square \, \Psi (\beta^a , \nu^{\alpha}) = 0 \, ,
\end{equation}
where $\square$ denotes the (formal) Laplace-Beltrami operator on the infinite-dimensional Lorentz-signature curved coset manifold $E_{10} ({\mathbb R}) / K(E_{10} ({\mathbb R}))$.

As recently emphasized \cite{DNGRF07}, the gravity/coset correspondence sketched above suggests a new physical picture of the fate of space at a cosmological singularity. It suggests that, upon approaching a spacelike singularity, the description in terms of a spatial continuum breaks down and should be replaced by a purely abstract Lie algebraic description. In other words, one is led to the conclusion that space actually ``disappears'' (or ``de-emerges'') as the singularity is approached\footnote{We have in mind here a ``big crunch'', i.e. we conventionally consider that we are tending {\it toward} the singularity. {\it Mutatis mutandis}, we would say that space ``appears'' or ``emerges'' at a big bang.}. The gravity/coset duality suggests that there is no (quantum) ``bounce'' from an incoming collapsing universe to some outgoing expanding one \cite{hep-th/0504153}. Rather, it is suggested that ``life continues'' for an infinite ``affine time'' at a singularity, with the double understanding that: (i) life continues only in a totally new form (as in a kind of ``transmigration''), and (ii) an infinite affine time interval (measured, say, in the coordinate $t$ of Eq.~(\ref{eq3}) with a coset lapse function $n(t) = 1$) corresponds to a sub-Planckian interval of geometrical proper time\footnote{Indeed, it is found that the coset time $t$, with $n(t) = 1$, corresponds to a ``Zeno-like'' gravity coordinate time (with rescaled lapse $\tilde N = N / \sqrt g = 1$) which tends to $+\infty$ as the proper time tends to zero.}.

\end{document}